\documentclass[prb,twocolumn,superscriptaddress,showpacs,amsmath,amssymb]{revtex4}

\usepackage{graphicx}
\usepackage{dcolumn}
\usepackage{bm}

\begin{document}

\title{ Electrical Detection and Magnetic-Field Control of Spin States in Phosphorus-Doped Silicon}


\author{H.\,Morishita}
\affiliation{School of Fundamental Science and Technology, Keio University, Japan}

\author{L.\,S.\,Vlasenko}
\affiliation{A. F. Ioffe Physico-Technical Institute of Russian Academy of Sciences, Russia}

\author{H.\,Tanaka}
\author{K.\,Semba}
\affiliation{NTT Basic Research Laboratories, NTT Corporation, Japan}

\author{K.\,Sawano}
\author{Y.\,Shiraki}
\affiliation{Advanced Research Laboratories, Tokyo City University, Japan} 

\author{M.\,Eto}
\author{K.\,M.\,Itoh}
\email[]{kitoh@appi.keio.ac.jp}
\affiliation{School of Fundamental Science and Technology, Keio University, Japan}


\date{\today}

\begin{abstract}
Electron paramagnetic resonance of ensembles of phosphorus donors in silicon 
has been detected electrically with externally applied magnetic fields lower than 200 G.  
Because the spin Hamiltonian was dominated by the contact hyperfine term rather than by 
the Zeeman terms at such low magnetic fields, superposition states 
$ \alpha{}\left| \uparrow\downarrow \right>+\beta{}\left| \downarrow\uparrow \right>$ and
$-\beta{}\left| \uparrow\downarrow \right> + \alpha{}\left| \downarrow\uparrow \right>$ were formed  
between phosphorus electron and nuclear spins, and electron paramagnetic resonance
transitions between these superposition states and 
$\left| \uparrow\uparrow \right>$ or $\left| \downarrow\downarrow \right>$ states are observed clearly.  A continuous change of $\alpha{}$ and $\beta{}$ with the magnetic field was observed with a behavior fully 
consistent with theory of phosphorus donors in silicon.  
\end{abstract}


\maketitle



\section{Introduction}
A phosphorus in silicon is attracting much attention 
towards realization of solid-state quantum information processors. 
It can be viewed as a two-qubit system having one $^{31}$P nuclear spin ($I  = 1/2$) and 
one electron spin ($S = 1/2$). \cite{Kane,Kane2000,all-silicon,Sarovar,EDMRQC}  
Coherent manipulation of its electronic states, \cite{MarNP,MarPRL,GWMPRL08,FelarXiv} coherent transfer of states between electron 
and nuclear-spins, \cite{entofPinSi} and large hyperpolarization 
of nuclear spins \cite{BoehmePRL2009, YSSPRL2009} have been demonstrated recently. 
Thanks to phosphorus's long spin dephasing time, \cite{Tyryshkin,Abe} 
enrichment of silicon with nuclear spin-free $^{28}$Si\,
has suppressed the background isotope fluctuation significantly \cite{SiIsotope1,SiIsotope2} to make possible the optical detection of 
$^{31}$P nuclear spin states. \cite{PLE}
Because a standard electron 
paramagnetic resonance (EPR)  measurement requires at least 10$^{9}$ spins or more, 
much more sensitive electrical detection methods of phosphorus EPR have been 
attracting attention.  \cite{MarNP,MarPRL,GWMPRL08,FelarXiv,MMSPRL2008,Solomon,Stich:JAP77,Stich:d-a,Stich:APL68,
pointdefect,P50,HighEDMR,Honig57} 
The record so far reported is the detection of 
$\sim{}$50 phosphorus spin states \cite{P50} and 
extensive efforts are underway worldwide to 
detect single phosphorus spin states. \cite{YMM09} 
All of the previous phosphorus EPR studies have been performed in 
the ``high-magnetic-field regime," which can be defined as $B \gg{}$200 G for 
phosphorus in silicon as we demonstrate later,
and have observed two EPR allowed transitions. \cite{MarNP,MarPRL,GWMPRL08,FelarXiv,BoehmePRL2009,Solomon,P50,HighEDMR,Honig57} 
The present work reports electrically detected magnetic resonance (EDMR) of phosphorus spin states 
in silicon which shows five of six possible transitions expected for the phosphorus donors 
in silicon under the low magnetic field $B \leq{}$200 G. 
 
The spin Hamiltonian of an isolated phosphorus atom placed in 
an externally applied magnetic field $B$ is given by;
\begin{equation}
\displaystyle {\cal H}_{\rm{Si:P}}=g_{e}\mu{}_{B}BS_{z}-g_{n}\mu{}_{n}BI_{z}+a\bm{S}\cdot{}\bm{I},
\label{Si:PHamil}
\end{equation}
where $\bm{S}$ and $\bm{I}$ are electron and phosphorus nuclear spins,
respectively.  
The first, second, and third terms represent the electron Zeeman, nuclear Zeeman, and contact hyperfine interaction between phosphorus electron and nuclear spins, respectively.  
Here $g_{e}\mu{}_{B}/2\pi{}\hbar{}\approx{}28$ GHz/T and 
$g_{n}\mu{}_{n}/2\pi{}\hbar{}\approx{}17.2$ MHz/T are given by electron 
and nuclear $g$-factors $g_{e}\approx{}1.9985$\, and
$g_{n}\approx{}2.2632$, respectively. \cite{CE, NuclMom}
The hyperfine constant is $a/2\pi{}\hbar{}\approx{} 117.5$ MHz. \cite{Feher1959}
Eigenstates of this spin Hamiltonian are given by;
\begin{eqnarray}
\left| 1 \right> &=& \left| \uparrow\uparrow \right>, 
\label{eq:eigen1} \\
\left| 2 \right> &=& \alpha{}\left| \uparrow\downarrow \right> + \beta{}\left| \downarrow\uparrow \right>, 
\label{eq:eigen2}\\
\left| 3 \right> &=& -\beta{}\left| \uparrow\downarrow \right> + \alpha{}\left| \downarrow\uparrow \right>, 
\label{eq:eigen3}\\
\left| 4 \right> &=& \left| \downarrow\downarrow \right>, 
\label{eq:eigen4}
\end{eqnarray}
where $\alpha{}=\cos{}\frac{\eta{}}{2}$ and $\beta{}=\sin{}\frac{\eta{}}{2}$.
$\eta$ is the angle between externally applied magnetic field direction and actual electron and nuclear spins precession axis given by 
$\tan{}\eta{}=\frac{a}{g_{e}\mu{}_{B}B-g_{n}\mu{}_{n}B}$. \cite{P-pESR}
A magnetic quantum number $+\frac{1}{2}$ $\left( -\frac{1}{2}\right) $ is 
represented by $\uparrow$ ($\downarrow$) and an arrow on the left (right) 
in each ket represents the electron (nuclear) spin state. 
Fig. 1(a) shows the magnetic field dependence of the four states expected for phosphorus in silicon.  At the high magnetic fields ($B \gg {}200$ G) $\alpha{} \simeq{} 1$ and $\beta{}\simeq{}0$, i.e., the four states simply become $\left| \uparrow\uparrow \right>$, $\left| \uparrow\downarrow \right>$, $\left| \downarrow\uparrow \right>$, and $\left| \downarrow\downarrow \right>$.  The EPR allowed transitions in this regime are limited to two: 
$\left| \uparrow\uparrow \right>\Leftrightarrow{}\left| \downarrow\uparrow \right>$ and 
$\left| \uparrow\downarrow \right>\Leftrightarrow{}\left| \downarrow\downarrow \right>$. 
In the low-magnetic-field regime defined by $B\leq{}$200 G, the eigenstates of $\left| 2 \right>$ and $\left| 3 \right>$ change continuously because $\eta$, i.e., $\alpha$ and $\beta$, change significantly with $B$ as shown in Fig. 1(b).  Therefore, the degree of superposition between
$\left| \uparrow\downarrow \right>$ and $\left| \downarrow\uparrow \right>$ that determines the EPR allowed transitions also changes with $B$.  
For example, the transition 
$\left| 1 \right> \Leftrightarrow{} \left| 2 \right>$ at the high magnetic field corresponds 
to the nuclear magnetic resonance (NMR) $\left| \uparrow\uparrow \right>\Leftrightarrow{}\left| \uparrow\downarrow \right>$ and it cannot be observed as the EPR.  However, the same transition becomes EPR observable at the low magnetic field because the EPR allowed component $\left| \uparrow\uparrow\right>\Leftrightarrow{}\left| \downarrow\uparrow \right>$ emerges with $\beta$.  Note that transitions $\left| 2 \right>\Leftrightarrow{}\left| {}3 \right>$ and $\left| 1 \right>\Leftrightarrow{}\left| {}4 \right>$ are weak because they are allowed only in the second order. Nevertheless, two aspects of quantum 
control that cannot be realized in the high-magnetic-field regime is expected to become possible in the low-magnetic-field regime; 1) controlling  the ratio of 
$\alpha{}$ and $\beta{}$ to change the degree of superposition by the magnetic field  
and 2) changing the population of the four states by utilizing 
the six transitions that are made allowed. 
The present work demonstrates these properties experimentally using 
electrical detection of phosphorus EPR  
and develops a quantitative theoretical model to support our observation.
\begin{figure}
	\begin{center}
	\includegraphics[width=7cm,clip]{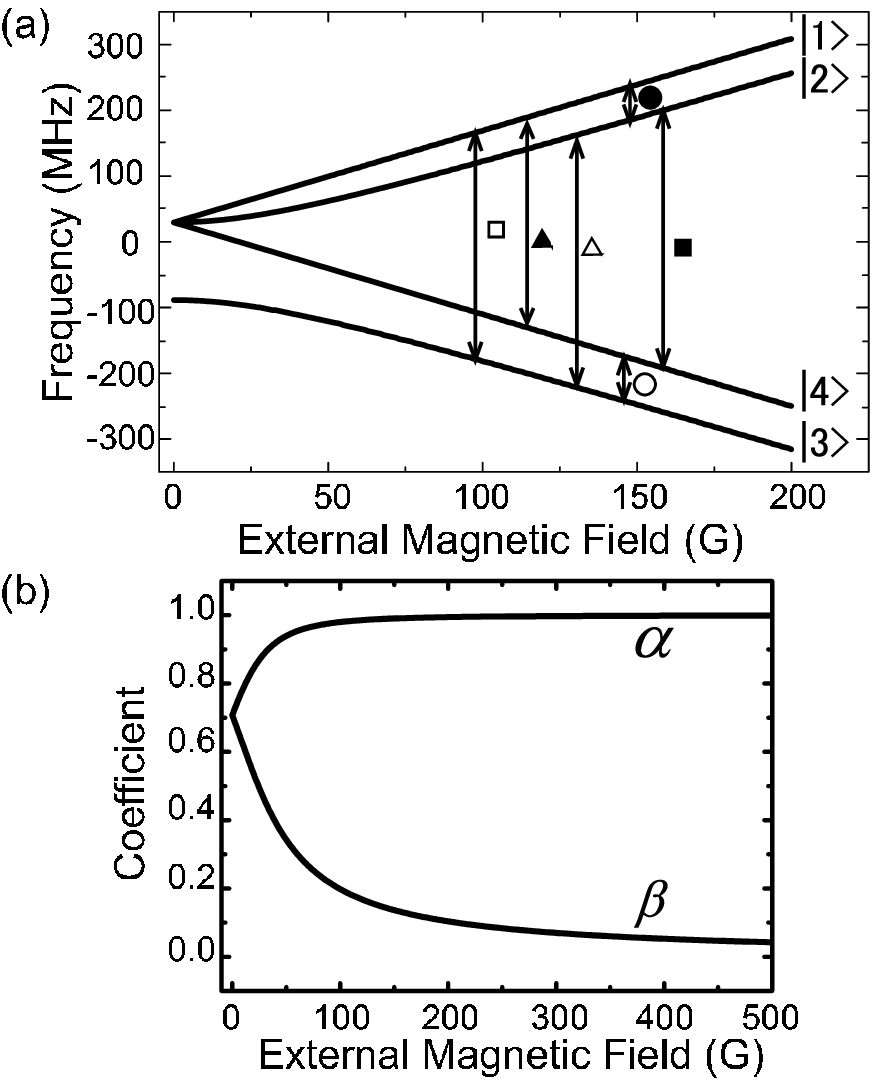}
	\caption{(a)Externally applied magnetic field dependence of the spin states of phosphorus in silicon
defined by 
Eqs. (\ref{eq:eigen1}) - (\ref{eq:eigen4}). 
Six allowed transitions 
$\left| 1 \right>\Leftrightarrow{} \left| 3 \right>$, 
$\left| 2 \right>\Leftrightarrow{}\left| 4 \right>$，
$\left| 2 \right>\Leftrightarrow{}\left| 3 \right>$, 
$\left| 1 \right>\Leftrightarrow{}\left| 4 \right>$,
$\left| 1 \right>\Leftrightarrow{}\left| 2 \right>$, and
$\left| 3 \right>\Leftrightarrow{}\left| 4 \right>$
are labeled 
by $\square{}$, $\blacksquare{}$, $\triangle{}$, $\blacktriangle{}$, $\bullet{}$, and $\circ{}$, respectively.
(b) Externally applied magnetic field dependence of $\alpha{}$ and $\beta{}$. $\alpha{}\rightarrow{}1$ and $\beta{} \rightarrow{} 0$ for $B\gg{}$200 G. }
	\label{fig:SiP}
	\end{center}
\end{figure}

\section{Experimental}
\begin{figure}
	\begin{center}
	\includegraphics[width=8cm,clip]{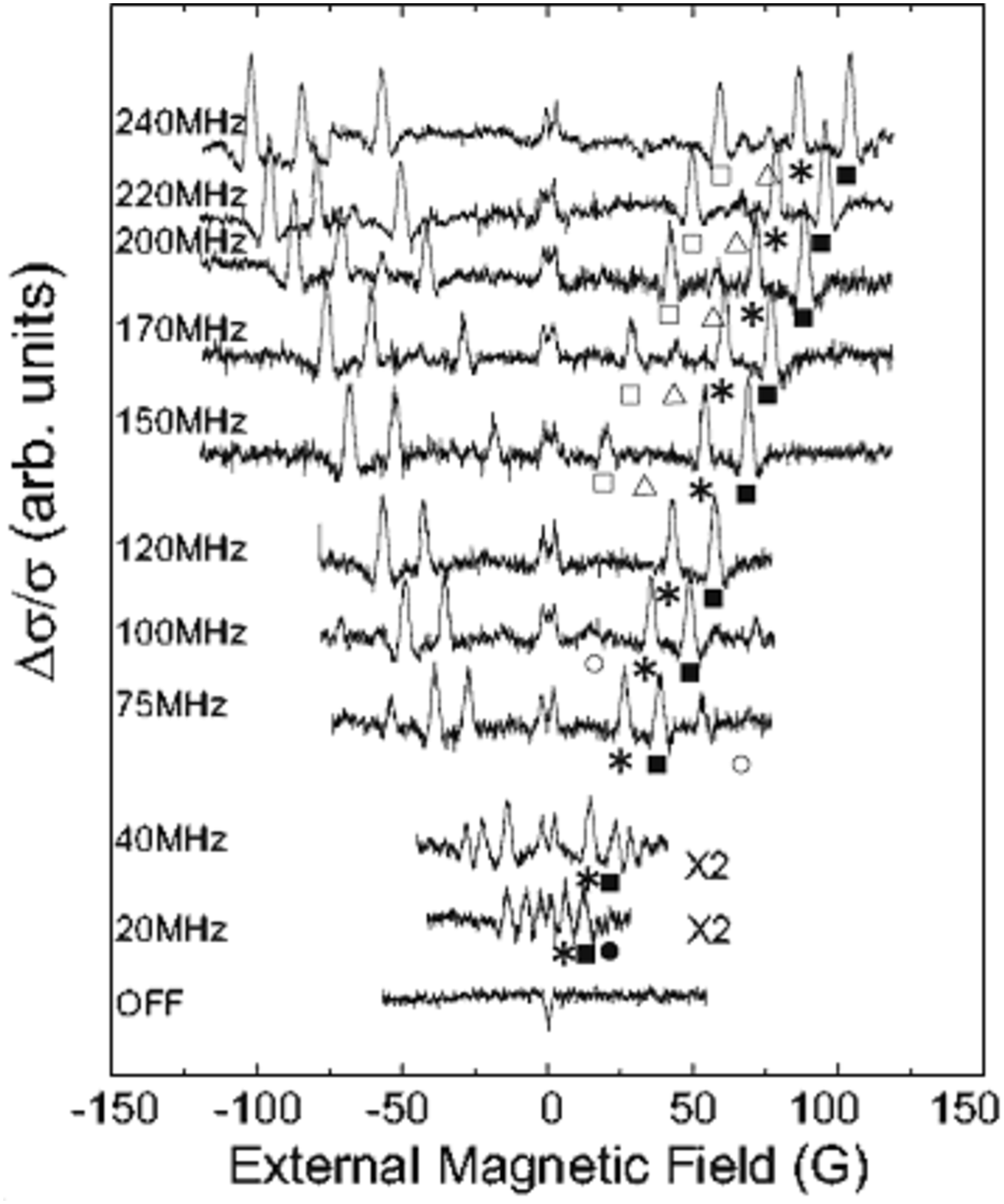}
	\caption{EDMR signals (change in the sample photoconductivity under continuous 
white light illumination from a halogen lamp) vs. 
externally applied magnetic field under irradiation of different 
500 mW radio frequencies (RF) as indicated in the figure. The sample is 
phosphorus-doped bulk silicon single crystal ([P]$\sim{}$10$^{16}$ cm$^{-3}$) 
kept at $T$ = 5 K during the measurement. The peaks indicated 
by $\square{}$ correspond to the transition
$\left| 1 \right>\Leftrightarrow{} \left| 3 \right>$ labeled by the same mark in Fig. \ref{fig:SiP}(a).  
Likewise, $\blacksquare{}$, $\triangle{}$, $\bullet{}$, and $\circ{}$
correspond to 
$\left| 2 \right>\Leftrightarrow{}\left| 4 \right>$，
$\left| 2 \right>\Leftrightarrow{}\left| 3 \right>$,
$\left| 1 \right>\Leftrightarrow{}\left| 2 \right>$, and
$\left| 3 \right>\Leftrightarrow{}\left| 4 \right>$,
respectively.  
Peaks labeled by $\ast{}$ correspond predominantly to the paramagnetic resonance of the 
interface center for the reason discussed in the text.
}
	\label{fig1}
	\end{center}
\end{figure}

A sample was a bulk Czochralski-grown n-type 
silicon single crystal having phosphorus concentration 
$\sim{}$10$^{16}$ cm$^{-3}$. It was cut into a rectangular 
shape of the dimension 8$\times$2$\times$1 mm$^3$. 
Ohmic contacts were prepared at both ends of 
the long axis by arsenic implantation of 
$2\times{}10^{15}\,\rm{cm}^{-2}$ at 25 keV followed by 
annealing at 980 $^{\circ }$C for 25 seconds and 
vacuum deposition of the 5-nm-thick palladium and 50-nm-thick gold layers.
The sample was placed in a cryostat with optical windows.   
A white light from a halogen lamp placed outside of 
the cryostat was focused onto the sample through the optical window for steady state excitation 
of the electron-hole pairs to maintain the sample resistance at $\sim$10 k$\Omega{}$.  
The sample was connected with a series resistor of 10 k$\Omega{}$.
A constant voltage of typically 10 V was applied to the series of 
the sample and resistor.
A coaxial cable was used to connect a RF source 
with an irradiation coil whose opposite side
was connected to a 50 $\Omega{}$ terminator.  Externally applied magnetic 
field was provided by a 300 mm bore electrical 
magnet. Another pair coil was placed in the cryostat to modulate 
the externally applied magnetic field for the lock-in detection of the 
divider voltage corresponding to the change in the sample 
photoconductivity (EDMR signal). 

\section{Results and Discussions}

\subsection{EDMR peak positions}

Fig. \ref{fig1} shows the EDMR signals obtained at $T$ = 5 K.
By changing the irradiation frequencies, six different transition 
peaks labeled $\square{}$, $\blacksquare{}$, $\triangle{}$, 
$\bullet{}$, $\circ{}$, and $\ast{}$ are observed clearly.

\begin{figure}[]
	\begin{center}
	\includegraphics[width=8cm,clip]{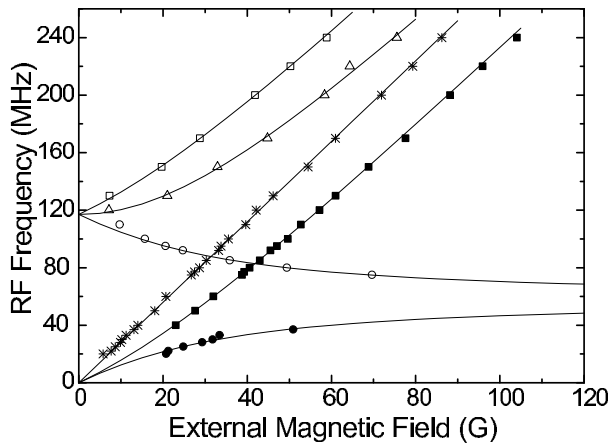}
	\caption{
A plot of irradiated frequency vs. externally applied magnetic field 
showing positions of experimentally 
determined peaks represented by the same marks as in Figs. \ref{fig:SiP} and \ref{fig1}.
The marks indicated by $\square{}$, $\blacksquare{}$, $\triangle{}$, $\bullet{}$, $\circ{}$, and $\ast{}$ correspond to 
$\left| 1 \right>\Leftrightarrow{} \left| 3 \right>$, 
$\left| 2 \right>\Leftrightarrow{}\left| 4 \right>$，
$\left| 2 \right>\Leftrightarrow{}\left| 3 \right>$,
$\left| 1 \right>\Leftrightarrow{}\left| 2 \right>$,
$\left| 3 \right>\Leftrightarrow{}\left| 4 \right>$,
and interface center transitions, respectively.  
Solid curves are rigorous theoretical calculations of resonance position of phosphorus in silicon, 
which show excellent agreement with experiments．}
	\label{fig2}
	\end{center}
\end{figure}

\begin{figure}[h]
	\begin{center}
	\includegraphics[width=8cm,clip]{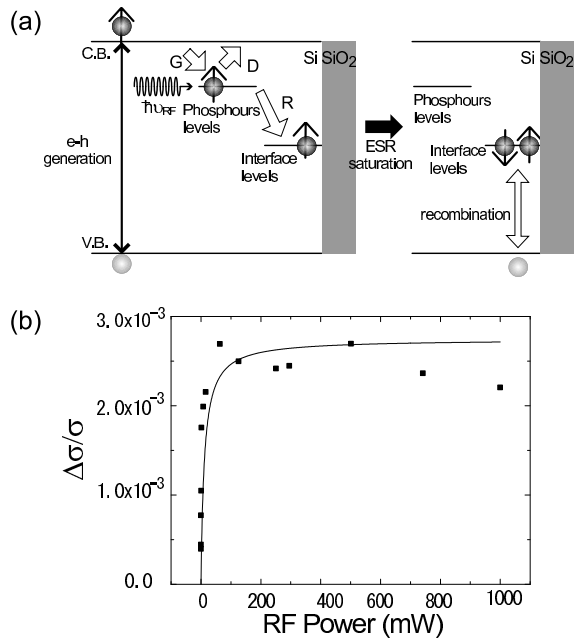}
	\caption{
(a) A schematic diagram of the EDMR mechanism.  An electron at the 
phosphorus level undergoes spin resonance and falls to 
the interface level when  the spin direction of phosphorus state and that of interface state 
form a spin singlet. Once the electron 
bound to phosphorus is gone, the phosphorus capturers another electron
from the conduction band, leading to the change in the photoconductivity.
Here the conduction electrons are captured by phosphorus at the rate 
$G$ and electrons at phosphorus go back to the conduction band at the 
rate $D$ or are captured by the interface states at the rate $R$. 
(b) RF power dependence of the EDMR signal for the transition 
$\left| 2 \right>\Leftrightarrow{}\left| 4 \right>$. The solid curve is 
the fitting using Eqs. (\ref{eq:late2}) and (\ref{eq:late3}).
}
	\label{fig3}
	\end{center}
\end{figure}

Fig. \ref{fig2} shows externally applied magnetic field 
vs. RF frequencies of the six observed resonance positions. 
Solid curves are theoretically expected results for different 
transitions of phosphorus in silicon using Eqs. (\ref{eq:eigen1}) - (\ref{eq:eigen4}) that are calculated rigorously with no fitting 
parameter. The excellent quantitative agreement 
between our experiment and theory supports the successful observation 
of the transitions listed in the captions of 
Figs. \ref{fig:SiP}, \ref{fig1}, and \ref{fig2}.

The transition indicated by $\ast{}$ agrees very well with the theoretically predicted resonance positions for 
$\left| 1 \right>\Leftrightarrow{}\left| 4 \right>$ labeled by $\blacktriangle{}$. However, we conclude that this peak is composed predominantly by the EPR transition of the paramagnetic defects situating around the interface between bulk Si and native surface oxide SiO$_2$ for the following reasons.  As apparent from Fig. \ref{fig:SiP}(a), the transition $\left| 1 \right>\Leftrightarrow{}\left| 4 \right>$ indicated by $\blacktriangle{}$  corresponds to a typical electron Zeeman transition whose energy is approximately proportional to $B$.  Therefore, EPR transition energies of other defects with very little hyperfine interaction, such as those of the interface spin states, \cite{Lepine, Pb} overlap with this transition.  Here the EPR of the interface spin state is allowed in the first order and, therefore, expected to have strong intensity.  
However, the transition $\left| 1 \right>\Leftrightarrow{}\left| 4 \right>$ labeled by $\blacktriangle{}$ is allowed only in the second order and should demonstrate approximately the same intensity as the transition $\left| 2 \right>\Leftrightarrow{}\left| 3 \right>$ labeled by $\triangle{}$ in Fig. 2.  The fact that the intensity of $\ast{}$ being much stronger than that of $\triangle{}$ suggests that the $\ast{}$ transition arises predominantly from the EPR transition of the interface defects that is allowed in the first order.  The importance of the presence of the interface defects was confirmed by removing the surface oxide using a dilute hydrofluoric (HF) solution.  The HF treatment made all of the EDMR signals nearly invisible. 
Leaving the sample for a few days in air to cover the sample surface with native oxide again retrieved the intensity of the all EDMR signals completely. 

\subsection{RF power and magnetic field dependencies of the phosphorus EDMR }
The fact that the presence of the surface oxide is 
needed to observe the EDMR signals allows us to 
develop a theoretical model describing the RF power 
and magnetic field dependencies of the EDMR signal 
intensity as the following.
We assume that the interface defects act as spin-dependent recombination centers for electrons 
bound to phosphorus donors.  In this sense, our model is an extension 
of existing two-level spin-dependent recombination model \cite{MarNP,MarPRL,GWMPRL08,FelarXiv, Stich:JAP77,Stich:d-a, pointdefect, White77, KSM,Haberkorn80, L'vov82, Vlasenko95,compKSM,Barabanov98, exchangeofEDMR} to the four levels.  
The spin Hamiltonian of the system during EDMR is given by
\begin{equation}
\displaystyle {\cal H}_{\rm{EDMR}}=g_{e}\mu{}_{B}B_0S_z+
 g_{e}\mu{}_{B}B_1S_{x}\cos{}(\omega t)+
 a\bm{S}\cdot{}\bm{I}.
\label{eq:system}
\end{equation}
Here the nuclear Zeeman term is neglected 
because $g_{n}\mu{}_{n}$ is approximately $10^{3}$ times 
smaller than $g_{e}\mu{}_{B}$ and the $J$ coupling term 
($J\bm{S}\cdot{}\bm{S}_{1}$ where 
$\bm{S}$ and $\bm{S}_{1}$ are the electron spins of the phosphorus
and interface states, respectively) is also neglected assuming $a\gg{}J$.
The second term arises from the RF irradiation 
and this perturbation term is defined as $\displaystyle {\cal H'} $  to calculate 
the transition probability $W$ for the system described by Eq. (\ref{eq:system}) using the Fermi's golden rule, 
\begin{equation}
W = \frac{2\pi{}}{\hbar{}}\left| \left< f \right| \displaystyle {\cal H'} \left| i \right> \right|^{2}
\delta{}(E_{f}-E_{i}-\hbar{}\nu{}).
\label{eq:fermi}
\end{equation}
The electron-hole recombination via interface states takes place only 
when $\bm{S}$ and $\bm{S}_{1}$ form a spin singlet and 
does not occur when they form a spin triplet to establish a ``spin blockade." \cite{Ono} 

\begin{table}
\caption[]{Recombination rate of each transition between phosphorus and interface spin states. \cite{ex1}}
\begin{ruledtabular}
\begin{tabular}{cccc}
$R_{i\sigma}$ & Phosphorus & Interface & Recombination rate \\ \hline
$R_{1\uparrow{}}$ & $\left| \uparrow{}\uparrow{}\right>$ &  $\left| \uparrow{} \right>$ & 0 \\ 
$R_{1\downarrow{}}$ & $\left| \uparrow{}\uparrow{}\right>$ &  $\left| \downarrow{} \right>$ & $\frac{1}{2}R$ \\
$R_{2\uparrow{}}$ & $\alpha{}\left| \uparrow{}\downarrow{}\right>$+$\beta{}\left| \downarrow{}\uparrow{}\right>$ &  $\left| \uparrow{} \right>$ & $\frac{1}{2}\beta{}^{2}R$\\
$R_{2\downarrow{}}$ & $\alpha{}\left| \uparrow{}\downarrow{}\right>$+$\beta{}\left| \downarrow{}\uparrow{}\right>$ &  $\left| \downarrow{} \right>$ & $\frac{1}{2}\alpha{}^{2}R$\\
$R_{3\uparrow{}}$ & $-\beta{}\left| \uparrow{}\downarrow{}\right>$+$\alpha{}\left| \downarrow{}\uparrow{}\right>$ &  $\left| \uparrow{} \right>$ & $\frac{1}{2}\alpha{}^{2}R$\\
$R_{3\downarrow{}}$ & $-\beta{}\left| \uparrow{}\downarrow{}\right>$+$\alpha{}\left| \downarrow{}\uparrow{}\right>$ &  $\left| \downarrow{} \right>$ & $\frac{1}{2}\beta{}^{2}R$\\
$R_{4\uparrow{}}$ & $\left| \downarrow{}\downarrow{}\right>$ &  $\left| \uparrow{} \right>$ & $\frac{1}{2}R$\\
$R_{4\downarrow{}}$ & $\left| \downarrow{}\downarrow{}\right>$ &  $\left| \downarrow{} \right>$ & 0\\
\end{tabular}
\label{tab1}
\end{ruledtabular}
\end{table}

Other important essences of our model are described in the 
caption of Fig. \ref{fig3}(a). Let us consider $\left| i \right>$ 
where $i=1,\, 2,\, 3,\,$ or $4$ is one of the four phosphorus spin states as
defined by Eqs. (\ref{eq:eigen1}) - (\ref{eq:eigen4}) and 
$\left| \sigma \right>$ where $\sigma{}=\,\uparrow{}$ or $\downarrow{}$ corresponds to spin up or down of the interface state, respectively.
Using $G$, $D$, and $R$ defined in Fig. \ref{fig3}(a), 
we obtain the rate equation
\begin{equation}
\frac{d}{dt}N_{i\sigma{}} = G(N-\sum_{j\sigma{}'}N_{j\sigma{}'})-(D+R_{i\sigma{}})N_{i\sigma{}},
\end{equation}
where $N$ is the total number of electron pairs and 
$N_{i\sigma{}}$ is the number of electron pairs in spin states
$i$ and $\sigma$.  
The recombination rates $R_{i\sigma{}}$ that have been obtained using the method described in Ref. 43 are listed in Table \ref{tab1}. 
The pairs with ($i, \sigma{}$) with $R_{i\sigma{}}=0$ correspond to the spin blockade. 
Now we consider a representing example where the irradiated  
RF is in resonance with the $\left| 2 \right>\Leftrightarrow{}\left| 4 \right>$ transition.  In this case; 
\begin{eqnarray}
\frac{d}{dt}N_{2\sigma{}} = G\left( N-\sum_{j\sigma{}'}N_{j\sigma{}'}\right) -\left( D+R_{2\sigma{}}\right) N_{2\sigma{}} \notag{} \\
-W\left( N_{2\sigma{}}-N_{4\sigma{}}\right), \hspace{2.6cm} \\
\frac{d}{dt}N_{4\sigma{}} = G\left( N-\sum_{j\sigma{}'}N_{j\sigma{}'}\right) -\left( D+R_{4\sigma{}}\right) N_{4\sigma{}}\notag{} \\
+W\left( N_{2\sigma{}}-N_{4\sigma{}}\right), \hspace{2.6cm}
\end{eqnarray}
where $W \propto{}\left( g_{e}\mu{}_{B}B_{1}/2\right) ^{2}\alpha{}^{2}$ is the transition probability 
that is proportional to the RF irradiation power around the origin. \cite{ex3}
This partly lifts the spin blockade and promotes the recombination. 
The steady state solution for the number of electron recombining $I(W)$ 
becomes;
\begin{eqnarray}
I(W) = \frac{NG}{1+G\left[ \sum_{i\sigma{}}\frac{1}{D+R_{i\sigma{}}}-WX(W)\right]} \hspace{1cm}\notag{}\\
\times{}\left[ \sum_{i\sigma{}}\frac{R_{i\sigma{}}}{D+R_{i\sigma{}}}+WDX(W) \right], 
\label{eq:late1}\\
X(W) = \sum_{\sigma{}}\frac{1}{\left( D+R_{2\sigma{}} \right)\left( D+R_{4\sigma{}} \right)}\hspace{2cm} \notag{} \\
\times{}\frac{\left( R_{2\sigma{}}-R_{4\sigma{}}\right)^{2}}{\left( 2D+R_{2\sigma{}}+R_{4\sigma{}} \right) W +\left( D+R_{2\sigma{}} \right)\left( D+R_{4\sigma{}} \right) }.
\label{eq:late2}
\end{eqnarray}
Then the EDMR signal intensity $S_{EDMR}$ is obtained as; 
\begin{eqnarray} 
S_{EDMR} = I(W)- I(0)\hspace{3.4cm}  \notag{}\\
= NG\frac{D+8G}{1+G\sum_{i\sigma{}}\frac{1}{D+R_{i\sigma{}}}} \hspace{2cm} \notag{}\\
\times{}\frac{WX(W)}{1+G\left[ \sum_{i\sigma{}}\frac{1}{D+R_{i\sigma{}}}-WX(W)\right] }.
\label{eq:late3}
\end{eqnarray}
\begin{figure}
	\begin{center}
	\includegraphics[width=8cm,clip]{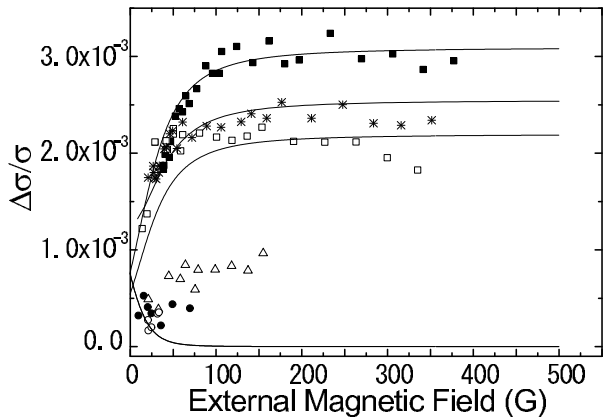}
	\caption{
EDMR intensity vs. externally applied magnetic field.
Experimentally 
determined positions are represented by the same marks in as Fig. \ref{fig1}.
The peaks indicated by $\square{}$, $\blacksquare{}$, $\triangle{}$, $\bullet{}$, $\circ{}$, and $\ast{}$ correspond to 
$\left| 1 \right>\Leftrightarrow{} \left| 3 \right>$, 
$\left| 2 \right>\Leftrightarrow{}\left| 4 \right>$，
$\left| 2 \right>\Leftrightarrow{}\left| 3 \right>$,
$\left| 1 \right>\Leftrightarrow{}\left| 2 \right>$,
$\left| 3 \right>\Leftrightarrow{}\left| 4 \right>$,
and interface center transitions, respectively.  
Solid curves are fittings for allowed transitions 
$\square{}$, $\blacksquare{}$, $\bullet{}$, and $\circ{}$
using Eqs. (\ref{eq:late3}) and (\ref{eq:late4}) with 
$WX(W)$ as described in the text.
The fits for $\bullet{}$ and $\circ{}$ overlap completely.
The interface center transition $\ast{}$ is fitted 
with Eqs. (\ref{eq:late3}) and (\ref{eq:interface2}).} 
	\label{fig4}
	\end{center}
\end{figure}
This result shows that the signal intensity is proportional to 
the irradiated RF power and, therefore, to $W$ around the origin. 
This corresponds to our 
experimental observation shown in Fig. \ref{fig3}(b) when 
$\Delta{}\sigma{} / \sigma{}$ is defined as $S_{EDMR} / I(0)$.  
A solid curve shown in the figure is the successful fitting 
by Eqs. (\ref{eq:late2}) and (\ref{eq:late3}) using $N$, $G$, $D$, and 
$R$ as fitting parameters．
A set of appropriate values we found are 
$N$ = 6.4$\times{}10^{5}$ cm$^{-3}$, $G$ = 5.0$\times{}10^{-9}$ sec$^{-1}$, 
$D$ = 2.3$\times{}10^{1}$ sec$^{-1}$, and $R$ = 1.7$\times{}10^{1}$ sec$^{-1}$. 

Finally, we show in Fig. \ref{fig4} the magnetic 
field dependence of the EDMR signal intensity.  We used
the RF power of 500 mW, which was large enough to saturate 
the signal as shown in Fig. \ref{fig3}(b).  Here $W \gg{} D, R$. 
Now the two $WX(W)$'s in Eq. (\ref{eq:late2}) are 
replaced by a constant;
\begin{equation}
WX(W)\rightarrow{}\sum_{\sigma{}}\frac{1}{\left( D+R_{2\sigma{}} \right)\left( D+R_{4\sigma{}} \right)}
\cdot{}\frac{\left( R_{2\sigma{}}-R_{4\sigma{}}\right)^{2}}{ 2D+R_{2\sigma{}}+R_{4\sigma{}} }.
\label{eq:late4}
\end{equation}
Such relations with appropriate $R_{i\sigma{}}$ have 
been used to fit representative experimental 
results shown in Fig. \ref{fig4}.
Note that our model is not applicable to second-order allowed transitions such as $\left| 2 \right>\Leftrightarrow{}\left| 3 \right>$ and $\left| 1 \right>\Leftrightarrow{}\left| 4 \right>$.
From the fitting of the first-order allowed $\left| 2 \right>\Leftrightarrow{}\left| 4 \right>$ transition, we obtain a set of appropriate values $N$ = 1.0$\times{}10^{4}$ cm$^{-3}$,
$G$ = 6.7$\times{}10^{-6}$ sec$^{-1}$, $D$ = 2.9$\times{}10^{1}$ sec$^{-1}$, and  
$R$ = 7.0$\times{}10^{-1}$ sec$^{-1}$.
Note that the magnetic field dependence of $\alpha$ and $\beta$ employed here is same as the one shown in Fig. \ref{fig:SiP}(b).  Therefore, the excellent agreement between our experiment and model shows that the coefficients of superposition can be controlled by the choice of $B$ and reach the maximally entangled states 
$1/\sqrt 2 \left( \left| \uparrow\downarrow \right>+\left| \downarrow\uparrow \right>\right)$
in the limit of $B$ = 0 as expected. \cite{Bennett96}

Similarly, we can derive a relation for 
the interface center EDMR, since the rotation of the interface electron spin $\bm{S}_{1}$ 
also lifts the spin blockade and enhances the recombination.  
We consider again a rate equation using $N$, $G$, $D$, and $R$ to obtain the same $S_{EDMR}$ as Eq. (\ref{eq:late3}) but different $X(W)$ from the phosphorus resonance case: \cite{ex2} 
\begin{eqnarray}
X(W)=\sum_{i=1}^{4}\frac{1}{\left( D+R_{i\uparrow{}}\right) \left( D+R_{i\downarrow{}} \right)}\hspace{2cm} \notag{} \\
\times{}\frac{\left( R_{i\uparrow{}}-R_{i\downarrow{}}\right)^2}
{\left( 2D+R_{i\uparrow{}}+R_{i\downarrow{}}\right) W+\left( D+R_{i\uparrow{}}\right) \left( D+R_{i\downarrow{}} \right)}.
\label{eq:interface}
\end{eqnarray}
When the signal intensity is saturated, $WX(W)$ is given by; 
\begin{equation}
WX(W)\rightarrow{}\sum_{i=1}^{4}\frac{1}{(D+R_{i\uparrow{}})(D+R_{i\downarrow{}})}\cdot{}\frac{(R_{i\uparrow{}}-R_{i\downarrow{}})^2}{2D+R_{i\uparrow{}}+R_{i\downarrow{}}}.
\label{eq:interface2}
\end{equation}
This result has been used to fit the behavior of the interface center 
peak ($\ast{}$)
in Fig. \ref{fig4}. 
A set of appropriate values we found are  
$N$ = 8$\times{}10^{3}$ cm$^{-3}$, $G$ = 2.0$\times{}10^{-5}$ sec$^{-1}$, 
$D$ = 3.6$\times{}10^{1}$ sec$^{-1}$, and $R$ = 3.9$\times{}10^{-1}$ sec$^{-1}$.
Here the values of $N$, $D$, and $R$ are approximately the same as the ones obtained for phosphorus 
but the value of $G$ is different.

A significance of the results shown in Fig. \ref{fig4} is 
that the intensity of transitions changes below 
200 G in accordance with theory.  The standard EPR allows for observation of only
$\left| 1 \right>\Leftrightarrow{}\left| 3 \right>$ and 
$\left| 2 \right>\Leftrightarrow{}\left| 4 \right>$ 
because $\beta = 0$ in the high magnetic fields.  
However, the  value of $\beta$ increases with decreasing the magnetic  
field, especially for the magnetic fields below 200 G, and approaches 1/$\sqrt 2$ as 
$B$ $\rightarrow$ 0.  Naturally, the intensity of 
$\left| 1 \right>\Leftrightarrow{}\left| 3 \right>$ and 
$\left| 2 \right>\Leftrightarrow{}\left| 4 \right>$
decreases because the components of the EPR allowed 
$\left| \uparrow\uparrow \right>\Leftrightarrow{}\left| \downarrow\uparrow \right>$
and 
$\left| \uparrow\downarrow\right>\Leftrightarrow{}\left| \downarrow\downarrow \right>$
diminish.
For the same reason transitions such as $\left| 1 \right>\Leftrightarrow{}\left| 2 \right>$ and 
$\left| 3 \right>\Leftrightarrow{}\left| 4 \right>$ appear 
only when $B$ $\leq{}$200 G.
This observation leads us to conclude that it is possible 
to form the superposition states between electron and 
nuclear spins of phosphorus in the regime of the low magnetic field $B$ $\leq{}$200 G.  Their superposition coefficients $\alpha$ and $\beta$ can be controlled 
simply by selecting an appropriate magnetic field.

\section{Conclusion}
Electron paramagnetic resonance spectroscopy of an ensemble of phosphorus donors in silicon has been performed successfully to map out the behavior of phosphorus spin states at the magnetic field lower than 200 G.  
Formation of the superposition states 
$ \alpha{}\left| \uparrow\downarrow \right>+\beta{}\left| \downarrow\uparrow \right>$ and
$-\beta{}\left| \uparrow\downarrow \right> + \alpha{}\left| \downarrow\uparrow \right>$ has been confirmed with the values of 
$\alpha{}$ and $\beta{}$ changing continuously with the magnetic field in accordance with theory of phosphorus in silicon.  Dependencies of the EDMR signal intensity on the RF power and magnetic field have been described successfully by a model assuming a spin-dependent recombination of phosphorus electrons via defects situating around the oxide/silicon interface.

\section{Acknowledgment}
We thank Martin Brandt for fruitful discussion. This work was supported in part by a Grant-in-Aid for Scientific Research by MEXT Specially Promoted Research $\sharp$18001002, in part by Special Coordination Funds for Promoting Science and Technology, in part by the JST-DFG Strategic Cooperative Program on Nanoelectronics, in part by the Strategic Information
and Communications R\&{}D Promotion Program (SCOPE) from the
Ministry of Internal Affairs and Communications of Japan
and in part by a Grant-in-Aid for the Global Center of Excellence at Keio University.


\begin{thebibliography}{}
\bibitem{Kane}
B. E. Kane, Nature \textbf{393}, 133 (1998).
\bibitem{Kane2000}
B. E. Kane, Fort. Physik \textbf{48}, 1023 (2000).
\bibitem{all-silicon}
K. M. Itoh, Solid State Commun. \textbf{133}, 747 (2005).
\bibitem{Sarovar}
M. Sarovar, K. C. Young, T. Schenkel, and K. B. Whaley, Phys. Rev. B \textbf{78}, 245302 (2008).
\bibitem{EDMRQC}
C. Boehme and K. Lips, Phys. Stat. Sol. (b) \textbf{233}, 427 (2002).
\bibitem{MarNP}
A. R. Stegner, C. Boehme, H. Huebl, M. Stutzmann, K. Lips, and M. S. Brandt, Nature Phys. \textbf{2}, 835 (2006).
\bibitem{MarPRL}
H. Huebl, F. Hoehne, B. Grolik, A. R. Stegner, M. Stutzmann, and M. S. Brandt, Phys. Rev. Lett. \textbf{100}, 177602 (2008).
\bibitem{GWMPRL08}
G. W. Morley, D. R. McCamey, H. A. Seipel, L.-C. Brunel, J. van Tol, and C. Boehme, Phys. Rev. Lett. \textbf{101}, 207602 (2008).
\bibitem{FelarXiv}
F. Hoehne, H. Huebl, B. Galler, M. Stutzmann, and M. S. Brandt, arXiv:0908.3612 (2009).
\bibitem{entofPinSi}
J. J. L. Morton, A. M. Tyryshkin, R. M. Brown, S. Shankar, B. W.
Lovett, A. Ardavan, T. Schenkel, E. E. Haller, J. W. Ager, and S. A. Lyon, Nature \textbf{455}, 1085 (2008).
\bibitem{BoehmePRL2009}
D. R. McCamey, J. van Tol, G. W. Morley, and C. Boehme, Phys. Rev. Lett. \textbf{102}, 027601 (2009). 
\bibitem{YSSPRL2009}
A. Yang, M. Steger, T. Sekiguchi, M. L. W. Thewalt, T. D. Ladd, K. M. Itoh, H. Riemann, N. V. Abrosimov, P. 
Becker, and H.-J. Pohl, Phys. Rev. Lett. \textbf{102}, 257401 (2009).
\bibitem{Tyryshkin}
A. M. Tyryshkin, J. J. L. Morton, S. C. Benjamin,
A. Ardavan, G. A. D. Briggs, J. W. Ager, and S. A. Lyon, J. Phys. Comd. Matt. \textbf{18}, S783 (2006).
\bibitem{Abe}
E. Abe, K. M. Itoh, J. Isoya, and S. Yamasaki, Phys. Rev. B \textbf{70}, 033204 (2004).
\bibitem{SiIsotope1}
K. M. Itoh, J. Kato, F. Uemura, A. K. Kaliteyevskii, O. N. Godisov, G. G. Devyatych, A. D. Bulanov, A. V. Gusev, I. D. Kovalev, P. G. Sennikov, H.-J.  Pohl, N. V. Abrosimov, and H. Riemann, Jpn. J. Appl. Phys. \textbf{42}, 6248 (2003).
\bibitem{SiIsotope2}
K.Takyu, K. M. Itoh, K. Oka, N. Saito, and V. I. Ozhogin, Jpn. J. Appl. Phys. \textbf{38}, L1493 (1999).
\bibitem{PLE}
A. Yang, M. Steger, D. Karaiskaj, M. L. W. Thewalt, M. Cardona, K. M. Itoh, H. Riemann, N. V. Abrosimov, M. F. Churbanov, A. V. Gusev, A. D. Bulanov, A. K. Kaliteevskii, O. N. Godisov, P. Becker, H.-J. Pohl, J. W. Ager III, and E. E. Haller, Phys. Rev. Lett. \textbf{97}, 227401 (2006).
\bibitem{MMSPRL2008}
G. W. Morley, D. R. McCamey, H. A. Seipel, L.-C. Brunel, J. van Tol, and C. Boehme, Phys. Rev. Lett. \textbf{101}, 207602 (2008).
\bibitem{Solomon}
J. Schmidt and I. Solomon, Comptes Rendus de L'Acad$\acute{\rm{e}}$mie des Sciences B \textbf{263}, 169 (1966).
\bibitem{Stich:JAP77}
B. Stich, S. Greulich-Weber, and J.-M. Speath, J. Appl. Phys. \textbf{77}, 1546 (1995).
\bibitem{Stich:d-a}
S. Greulich-Weber, B. Stich, and J.-M. Spaeth, Materials Science Forum  \textbf{196-201}, 1509 (1995).
\bibitem{pointdefect}
J.-M. Spaeth and H. Overhof, \textit{Point Defects in Semiconductors and Insulators} (Springer, 2002), Chap. 7.
\bibitem{Stich:APL68}
B. Stich, S. Greulich-Weber, and J.-M. Speath, Appl. Phys. Lett. \textbf{68}, 1102 (1996).
\bibitem{P50}
D. R. McCamey, H. Huebl, M. S. Brandt, W. D. Hutchison, J. C. McCallum, R. G. Clark, and A. R. Hamilton, Appl. Phys. Lett. \textbf{89}, 182115 (2006).
\bibitem{HighEDMR}
D. R. McCamey, G. W. Morley, H. A. Seipel, L.-C. Brunel, J. van Tol, and C. Boehme, Phys. Rev. B \textbf{78}, 045303 (2008).
\bibitem{Honig57}
A. Honig and M. Moroz, Rev. Sci. Instrum. \textbf{49}, 183 (1957).
\bibitem{YMM09}
K. Y. Tan, K. W. Chan, M. M\"{o}tt\"{o}nen, A. Morello, C. Yang, J. van Donkelaar, A. Alves, J. M. Pirkkalainen, D. N. Jamieson, R. G. Clark, and A. S. Dzurak, arXiv:0905.4358v2 (2009).
\bibitem{CE}
C. F. Young, E. H. Poindexter, G. J. Gerardi, W. L. Warren, and D. J. Keeble, Phys. Rev. B \textbf{55}, 16245 (1997).
\bibitem{NuclMom}
J. E. Mack, Rev. Mod. Phys. \textbf{22}, 64 (1950).
\bibitem{Feher1959}
G. Feher, Phys. Rev. \textbf{114}, 1219 (1959).
\bibitem{P-pESR}
A. Schweiger and G. Jeschke, \textit{Principles of pulse Electron Paramagnetic Resonance} (OXFORD, 2001), Chap. 3, Sec. 3.5, pp.58 - 62.
\bibitem{Lepine}
D. J. L$\acute{\rm{e}}$pine, Phys. Rev. B \textbf{6}, 436 (1970).
\bibitem{Pb}
J. L. Cantin and H. J. von Bardeleben, J. Non-Cryst. Solids \textbf{303}, 175 (2002).
\bibitem{White77}
R. M. White and J. F. Gouyet, Phys. Rev. B \textbf{16}, 2596 (1977).
\bibitem{KSM}
D. Kaplam, I. Solomon, and N. F. Mott, Le J. de Phys. Lett. \textbf{39}, L51 (1978).
\bibitem{Haberkorn80}
R. Haberkorn and W. Dietz, Solid State Commun. \textbf{35}, 505 (1980).
\bibitem{L'vov82}
V. S. L'vov, L. S. Mima, and O. V. Tretyak, Sov. Phys. JETP \textbf{56}, 897 (1982).
\bibitem{Vlasenko95}
L. S. Vlasenko, Yu. V. Martynov, T. Gregorkiewicz, and C. A. J. Ammerlaan, Phys. Rev. B \textbf{52}, 1144 (1995).
\bibitem{compKSM}
A. V. Barabanov, O. V. Tretiak, and V. A. L'vov, Phys. Rev. B \textbf{54}, 2571 (1996).
\bibitem{Barabanov98}
A. V. Barabanov, V. A. Lvov, and O. V. Tretyak, Phys. Stat. Sol. (b) \textbf{207}, 419 (1998).
\bibitem{exchangeofEDMR}
A. Gliesche, C. Michel, V. Rajevac, K. Lips, S. D. Baranovskii, F. Gebhard, and C. Boehme, Phys. Rev. B \textbf{77}, 245206 (2008).
\bibitem{Ono}
A similar phenomenon was observed in transport through double quantum dots in K. Ono et al., Science {\bf 297}, 1313 (2002).
\bibitem{ex1}
$R$ is the transition rate of an electron from phosphorus to the interface levels when $\bm{S}$ and $\bm{S_{1}}$ form a spin singlet. The transition rate $\propto{} \exp\big[ -r(a_{B}^{*\,-1}+a_{i}^{-1})\big]$, where $r$ is the distance between a phosphorus and an interface state and $a_{B}^{*}\,(a_{i})$ is the effective Bohr radius of phosphorus (interface) state. It is averaged with the weight of $P(r) = -\frac{d}{dr}\exp(-4\pi{}r^{3}n_{D}/3)$, where $n_{D}$ is the donor concentration.
\bibitem{ex3}
On the other hand, the transition probabilities of  $\left| 1 \right>\Leftrightarrow{}\left| 2 \right>$, $\left| 1 \right>\Leftrightarrow{}\left| 3 \right>$, and $\left| 3 \right>\Leftrightarrow{}\left| 4 \right>$ are proportional to $\beta{}^{2},\, \alpha{}^{2}$, and $\beta{}^{2}$, respectively.
\bibitem{Bennett96}
C. H. Bennett, H. J. Bernstein, S. Popescu, and B. Schumacher, Phys. Rev. A \textbf{53}, 2046 (1996).
\bibitem{ex2}
The rate equations given by $\frac{d}{dt}N_{i,\sigma{}} = G(N-\sum_{j\sigma{}'}N_{j\sigma{}'})-(W+D+R_{i\sigma{}})N_{i\sigma{}}+WN_{i\bar{\sigma{}}}$, where $\bar{\sigma{}} = \uparrow{} (\downarrow)$ for $\sigma{} = \downarrow{} (\uparrow)$ and $W\propto{}(g\mu{}_{B}B/2)^{2} $.
\end{thebibliography}
\end{document}